# Failure mechanism of monolayer graphene under hypervelocity impact of spherical projectile

Kang Xia[1], Haifei Zhan*,[1], De'an Hu[2], and Yuantong Gu**,[1]

[1]*School of Chemistry, Physics and Mechanical Engineering, Queensland University of Technology (QUT), Brisbane QLD 4001, Australia*
[2]*State Key Laboratory of Advanced Design and Manufacturing for Vehicle Body, Hunan University, Changsha 410082, PR China*

**Abstract**

The excellent mechanical properties of graphene have enabled it as appealing candidate in the field of impact protection or protective shield. By considering a monolayer graphene membrane, in this work, we assessed its deformation mechanisms under hypervelocity impact (from 2 to 6 km/s), based on a serial of *in silico* studies. It is found that the cracks are formed preferentially in the zigzag directions which are consistent with that observed from tensile deformation. Specifically, the boundary condition is found to exert an obvious influence on the stress distribution and transmission during the impact process, which eventually influences the penetration energy and crack growth. For similar sample size, the circular shape graphene possesses the best impact resistance, followed by hexagonal graphene membrane. Moreover, it is found the failure shape of graphene membrane has a strong relationship with the initial kinetic energy of the projectile. The higher kinetic energy, the more number the cracks. This study provides a fundamental understanding of the deformation mechanisms of monolayer graphene under impact, which is crucial in order to facilitate their emerging future applications for impact protection, such as protective shield from orbital debris for spacecraft.

**Introduction**

Graphene has drawn significant interests from both engineering and scientific communities due to its record breaking properties,[1-3] for instance, Young's modulus reaches 1 TPa and intrinsic strength as high as 130 GPa.[2] These intriguing properties have triggered broad potential engineering applications, such as flexible electronics,[4] coatings (against corrosion),[5] reinforcements (for nanocomposites),[6] and biomedical applications.[7] To facilitate various applications, great efforts have been devoted to realize massive production of graphene.[8-10] Through roll-to-roll technique, researchers have reported the successful fabrication of 30-inch graphene films.[11] Combining with their low mass density (~ 2220 kg/m$^3$), graphene shows appealing potentials in the field of impact protection and appears as a new candidate for shielding materials, such as combat armour,[12] and protective shield from orbital debris for spacecraft.[13]

In this regard, experimental trials have been conducted in 2014, in which the mechanical behaviours of multilayer graphene (thickness from 10 to 100 nm) under high-strain-rate have been studied by using miniaturized ballistic tests.[14] It is found that the specific penetration energy for multilayer graphene is more than 10 times higher than that reported for the macroscopic steel sheets in literature (under the ballistic speed of 600 m/s). With this understanding, several modelling works have followed based on molecular dynamics (MD) simulations to probe the deformation process of graphene under high-speed impact loading. For example, Yoon et al[15] investigated the penetration energy change of graphene with the presence of defects by utilizing ReaxFF bond-order-based potential (with an impact velocity of 5 km/s). They found that the density of mono-vacancies greatly affect the specific penetration of graphene under impact loading. Haque et al[16] studied the deformation of graphene under a rigid fullerene ($C_{180}$) projectile with the impact velocity ranging from 3.5 to 7.5 km/s.

Theoretically, graphene appears as good candidate for impact protection due to its high in-plane Young's modulus and low mass density, which leads to a superior speed of sound ( ~ 22.2 km/s). Such high speed of sound could efficiently transmit the concentrated tensile stress (in the impact area), and thus rapidly delocalize the impact energy.[17] Therefore, to utilize graphene for impact protections, several questions prompt: how the stress transmits from the impact area? How the boundary will influence the stress transmission? And how the crack growth will be affected by the stress transmission? We note that although Haque's work[16] has investigated the wave propagation, they have used an ultra-small and rigid bullet ($C_{180}$, with a diameter of only 1.21 nm). It is seen that they could only reproduce the formation of three cracks after perforation, which is different from the experimental observation (with four, five and six cracks). Moreover, since the energy delocalization is realized through stress wave propagation, a comprehensive understanding about how the boundary will influence the wave transmission is a necessity, which however is lacking in previous studies.

Therefore, in this work, we will first discuss the stress distribution and transmission during impact by varying the size of the projectile and also the sample shape. Afterwards, emphasis will be put on the failure shape of the graphene membrane, by examining influence from the projectile velocity, the relationship between crack length and penetration energy, as well as the stress wave.

**Computational details**

The fracture behaviour of graphene under hypervelocity impact is acquired through a series of large-scale MD simulations using the open-source LAMMPS code.[18] A spherical diamond projectile is adopted to impact the monolayer graphene. Initially, we consider a square graphene membrane which is fixed in all lateral directions (see Figure 1), and the sample has zigzag edge in *x* direction (armchair edge in *y* direction). The projectile is located in the middle of the graphene sample, and its centre has an initial distance of ~ 78.5 Å with the graphene centre. To isolate the influence from thermal fluctuations, the initial system temperature was chosen as 1 K.

The widely used adaptive intermolecular reactive empirical bond order (AIREBO) potential was employed to describe the C-C atomic interactions[19,20] within graphene membrane. This potential has been shown to represent well the binding energy and elastic properties of carbon materials. The C-C cut-off distance was chosen as 2.0 Å to accurately capture the bond breaking during impact.[21-23] For the C-C atomic interactions in the diamond, we adopted the Tersoff potential, which has been shown to well represent the binding energy of diamond.[24] The interactions between projectile and graphene are described by Morse potential.[25] During the simulation, the graphene was first relaxed to a minimum energy state using the conjugate gradient algorithm. We then used the Nose–Hoover thermostat[26] to equilibrate the whole system at 1 K (NVT ensemble) for 2000 fs. A small time step of 0.5 fs was selected for the simulation, which has also been adopted for the ballistic resistance test for carbon nanotube.[27] The equations of motion are integrated with time using a Velocity Verlet algorithm[28] and no periodic boundary conditions have been applied. Afterwards, a high velocity is applied to the projectile to impact the graphene, from 2 to 6 km/s (which is in the same order as that of the orbital debris[29-31]). To better describe the heating of graphene induced by the impact, no thermostat is applied during the impact process.

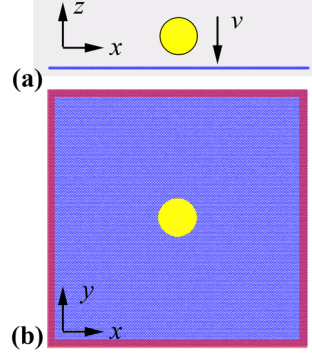

**Figure 1** Schematic view of the impact simulation setup. Upper is the front view and bottom is the top view. The red regions represent the fixed boundaries during impact.

During the simulation, the atomic stress in the graphene membrane is estimated based on the virial stress $\Pi^{\alpha\beta}$, which is expressed as[32]

$$\Pi^{\alpha\beta} = \frac{1}{\Omega}\sum_i \varpi_i \pi_i^{\alpha\beta}, \quad \pi_i^{\alpha\beta} = \frac{1}{\varpi_i}\left(-m_i v_i^\alpha v_i^\beta + \frac{1}{2}\sum_{j\neq i} F_{ij}^\alpha r_{ij}^\beta\right) \quad (1)$$

Here $\pi_i^{\alpha\beta}$ is the atomic stress associated with atom $i$. $\varpi_i$ is the effective volume of the $i$th atom and $\Omega$ is the volume of the whole system. $m_i$ and $v_i$ are the mass and velocity of the $i$th atom, respectively. $F_{ij}$ and $r_{ij}$ are the force and distance between atoms $i$ and $j$, respectively, and the indices $\alpha$ and $\beta$ denote the Cartesian components. In this work, the volume of the graphene membrane is estimated by assuming the graphene as a continuum media with a thickness of $h$ (which equals to the graphite interlayer distance 3.35 Å). Considering the stress states during impact, we tracked the atomic planar stress $\sigma_{xy}$ in the graphene membrane based on the atomic virial stress (Eq. 1), i.e., $\bar{\sigma}_{xy} = \bar{\sigma}_x + \bar{\sigma}_y$.

**Results and Discussions**

*Stress distribution and transmission*

Initially, we assess the stress distribution and also transmission during the impact process by considering a large square graphene membrane with a size of is $50 \times 50$ nm$^2$. The spherical projectile has a dimeter of 25 Å. A high velocity of 20 Å/ps (i.e., 2 km/s) is chosen to initiate the impact process, which is larger than the velocities used in experiments (< 1 km/s).

Figure 2a compares the total energy change of the graphene membrane ($\Delta E_{tot,G}$) and projectile ($\Delta E_{tot,P}$) during the impact process. Here, the total energy ($E_{tot}$) of the system is a sum of the kinetic energy ($E_{ke}$) and potential energy ($E_{pe}$). During the impact, the kinetic energy of projectile ($E_{ke,P}$) will decreases, and in the meanwhile its potential energy ($E_{pe,P}$) will increase due to the deformation (Figure 2b). Theoretically, the total energy change of projectile ($\Delta E_{tot,P}$) equals to that of the graphene membrane ($\Delta E_{tot,G}$), which is also the penetration energy ($E_p$). As is seen in Figure 2a, the profile of $\Delta E_{tot,G}$ almost overlaps with $\Delta E_{tot,P}$ during the penetration process. However, they deviate from each other after penetration. Specifically, the total energy change of projectile $\Delta E_{tot,P}$ remains as a constant after penetration, which is understandable as the simulation is under a vacuum condition. However, for graphene membrane, the energy change reaches a much larger value after penetration and then experiences a gradual increase before approaching a relatively flat profile (Figure 2a). Such obvious energy change deviation between graphene and projectile is supposed as stemmed from the fixed boundary, i.e., the final total energy change of graphene is a sum of the energy change of projectile and the work done by the fixed boundary. Despite such boundary effect (which will be revisited in the following), the overlapped profile of $\Delta E_{tot,G}$ and $\Delta E_{tot,P}$ during the penetration process signifies that the penetration energy ($E_p$) is equal to $\Delta E_{tot,P}$ (~ 2118 eV, the total energy loss of the projectile).

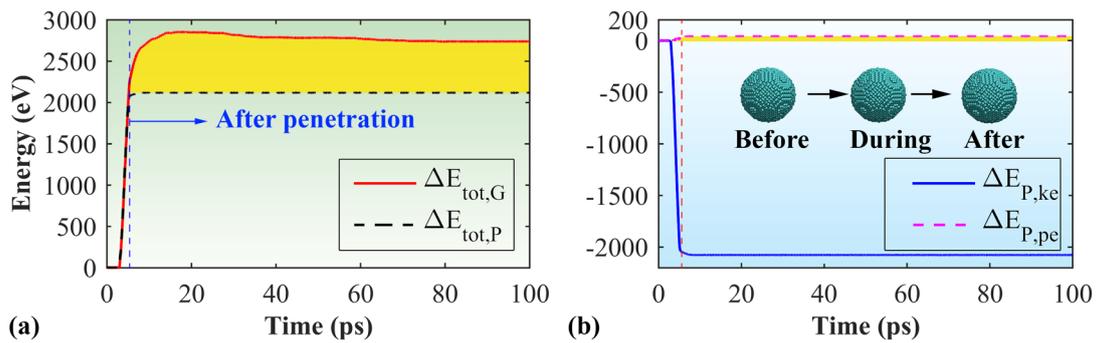

**Figure 2** (a) The total energy change of graphene and projectile during the impact process. (b) The kinetic energy change ($E_{ke,P}$) and potential energy change ($E_{pe,P}$) of the projectile. Insets show the morphology of the projectile before, during and after impact.

To exploit the deformation process, we examined the atomic configurations of the graphene membrane during the impact. We note that during the impact, although the potential energy of the projectile increases, the amount is ignorable (around ~ 42 eV) comparing with the change of its kinetic energy (about 2076 eV, Figure 2b). From the atomic configurations (insets of Figure 2b), no observable deformation is detected from the projectile during the whole impact process, therefore, following discussions will only focus on the graphene. As illustrated in Figure 3a, initial cracks are generated at the conical impact area (with a radius around 70 Å), which experiences a significant local deformation (~ 32 Å out-of-plane deformation) before the onset of crack propagations. As expected, the impact area suffers the highest planar stress, which decreases from the centre to the sample boundary. Here, the planar stress is the net virial stress along the $x$ and $y$-axis of the graphene. After the projectile completely perforated (or pass through) graphene, the high stress area decreases significantly (Figure 3b). Meanwhile, six initial cracks are found at the impact regime (with an angle around 60° between each other), and they behave differently from each other. Specifically, three of them stopped propagation quickly, while the other three (denoted as crack $c$, $a'$, and $c'$, respectively, in Figure 3c) continue propagating. The propagation distance of crack $a'$ is much smaller comparing with the other two, which is impeded by the two monoatomic linkages that connecting the dangling petals (inset of Figure 3c), and it stopped propagation at ~ 7.9 ps. As expected, the crack tips are the places where highest stress aggregate, and the regions aligned with the crack propagation direction (the dashed lines in Figure 3c) also experience higher stress. From Figure 3d, the final stage of the crack propagation happened at ~ 15.8 ps, featured by a crack kicking event in the zigzag direction. Such zigzag direction kicking phenomenon is also reported by Yin et al[33] in the final stage of uniaxial tensile loading simulation. More importantly, it is found that all crack edges are along zigzag direction (including the initial six cracks in Figure 3b), that is, the crack propagates preferentially along the zigzag direction, which is in line with previous studies.[33] The kicking crack growth stopped quickly, however, the stress at the tip region maintains higher than other parts (Figure 3d). With continuing simulation, the concentrated stress is fully released (Figure 3e), as well as the kinetic energy of the graphene petals gained from the projectile (Figure 3f).

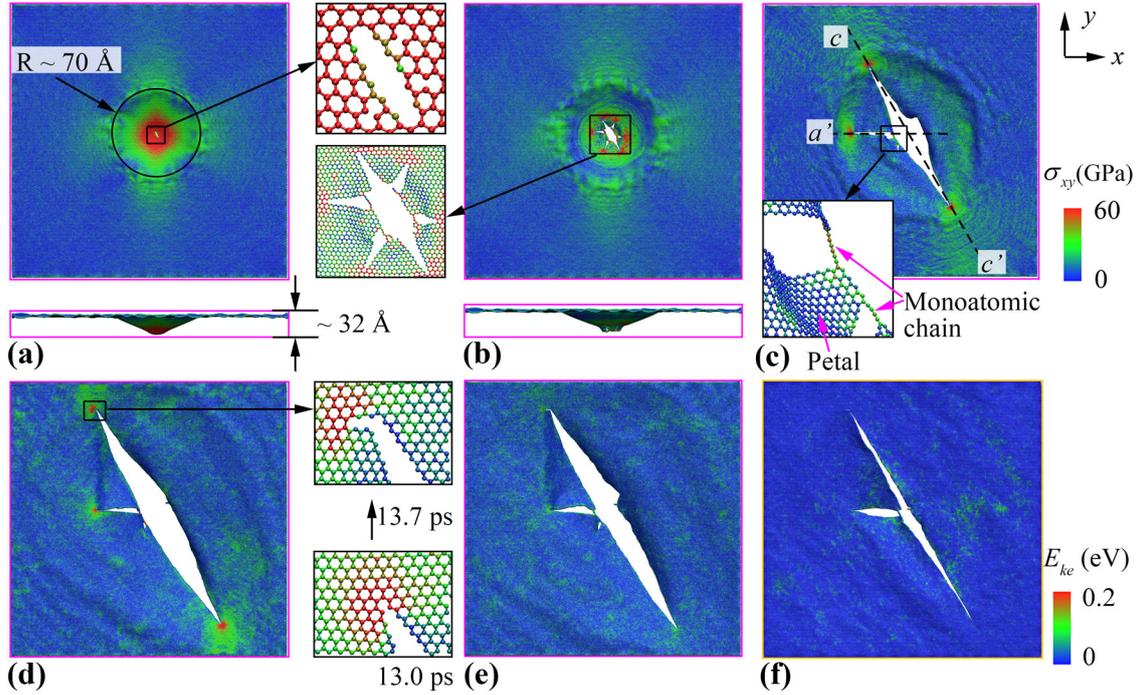

**Figure 3** Planar stress ($\sigma_{xy}$) distribution of the square graphene membrane at the simulation time of: (a) 5.1 ps, inset shows the formation of initial cracks at the impact area, (b) 5.5 ps, inset shows the initial cracks in six preferential directions, (c) 7.9 ps, inset shows the graphene petal being connected by monoatomic chains and the dashed lines illustrate the crack orientations, (d) 13.7 ps, inset shows the kicking phenomenon in zigzag direction, and (e) 27.6 ps, which shows the final failure shape of graphene membrane; (f) kinetic energy distribution at 27.6 ps. For all figures, atoms are coloured as red if their planar stress/kinetic energy is larger than the maximum value of the colour bar (i.e., 60 GPa for planar stress and 0.2 eV for kinetic energy). The projectile is not plotted in all figures.

Interestingly, we note that there are several clear "stress fringes" in Figure 3a before the crack propagation, which appear more concentrated along the directions that perpendicular to the four boundaries. Thus, to probe the potential influence from the boundaries on such "stress fringes" (or stress distribution) in graphene membrane, we consider another impact testing by adopting the same setting (e.g., same graphene size, projectile size and velocity, initial projectile position) but with only two boundaries being fixed (in *x* direction). Comparing with its counterpart with all fully fixed boundaries (Figure 3a), the graphene shows smaller penetration energy (~ 1777 eV) and the initial cracks appear much earlier (~ 4.6 ps). As shown in Figure 4a, the graphene shows smaller deformation region (with a radius of ~ 56 Å) and smaller out-of-plane deformation (~ 24 Å) when the crack appears. Such results indicate that the four boundaries fixed scenario is better in absorbing impact energy, with large deformation regions before the fracture starting to propagate. According to Figure 4b, the graphene also has six initial cracks, three of which have propagated with continuing simulation and form the final failure shape. We note that although the failure shape is different when the boundary conditions changed, the fracture mechanism are the same between Figure 3 and 4, i.e., zigzag-edged cracks, formation of monoatomic carbon chains, and stress concentration at the crack tips (Figure 4c).

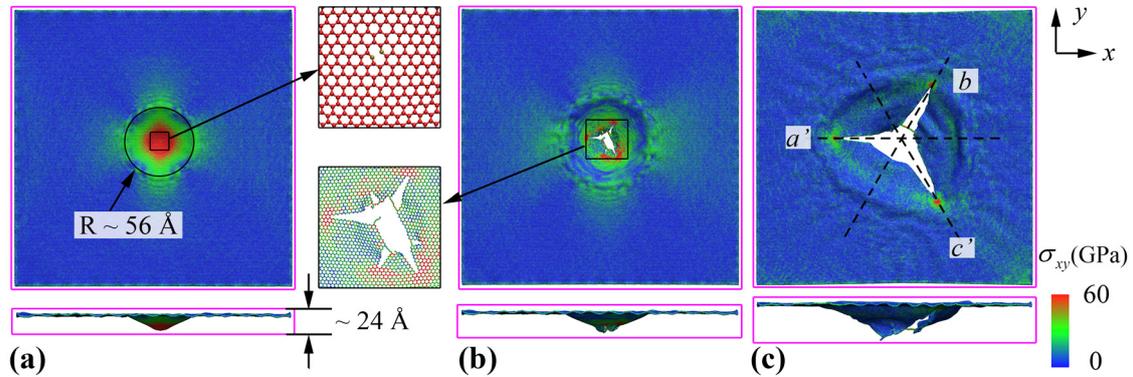

**Figure 4** Planar stress ($\sigma_{xy}$) distribution of the square graphene membrane with two fixed edges (in *x* direction) at the simulation time of: (a) 4.6 ps, inset shows the initial crack which is similar as that observed from its counterpart with all edges fixed in inset of Figure 3a, (b) 5.1 ps, inset shows the existences of monoatomic chains, and (c) 7.9 ps, which shows the final failure shape of graphene membrane, and the dashed lines illustrate the crack orientations.

Above results clearly show that the boundary will influence the stress distribution and transmission, which eventually leads to different failure shape of graphene membrane. To fully understand such influence, we apply the same simulation settings to the graphene membrane with different shapes (i.e., with different number of fixed edges). Considering the six preferential crack directions, a hexagonal graphene membrane is firstly assessed, which has a similar size as the above square sample with an edge length approximating 30 nm. Unlike the square graphene membrane, six clear stress fringes are formed perpendicular to the six edges, which have an average interval angle of about 60° (Figure 5a). Such scenario is also observed when we extend the hexagonal shape to a circular morphology (with a radius of 25 nm, Figure 5b). As illustrated in Figure 5c, unlike the sharp conical deformation shape observed from the square graphene membrane (Figure 3a and 4a), much larger area of the hexagonal graphene has involved with the deformation. With continuing simulation, the hexagonal graphene is found to reach a maximum out-of-plane deformation (about ~ 44 Å at 8.0 ps), and no initial crack or breaking of bonds is observed. Afterwards, the projectile is bounced back and the concentrated stress transmits to other deformable regions, leading to a relatively even stress distribution map (Figure 5d). Such deformation process is also observed from the circular graphene. These results imply that the fully fixed hexagonal or circular graphene membrane requires high penetration energy comparing with their square counterpart, i.e., more efficient for impact protection.

Of interest, we also compare the atomic planar stress distribution between the fully-fixed hexagonal and square graphene membrane. Here the stress distribution is calculated at selected simulation time. From Figure 5e, the hexagonal graphene possesses higher percentage of atoms with larger stress at 4.9 ps comparing with that of the square graphene (before cracks initiate). With further simulation, an evident gap is found between them. For the square graphene, the amount of atoms that experiencing higher stress increases due to the occurrence of penetration. According to the atomic configurations (Figure 3b, 3c and 3d), these atoms are actually located in the crack tips. In comparison, a relatively small increase of the number of high stress atoms are observed for the hexagonal graphene, which agrees with the atomic configurations (that no penetration is occurred). Overall, it is found that the boundaries would change the tensile stress distribution and also transmission, which will eventually influence the energy delocalization during impact. At the same projectile size and velocity, the fully-fixed hexagonal and circular graphene is more efficient for impact protection. We note that although current results are found in monolayer graphene membrane, a similar phenomenon is also expected from multi-layer graphene membrane.

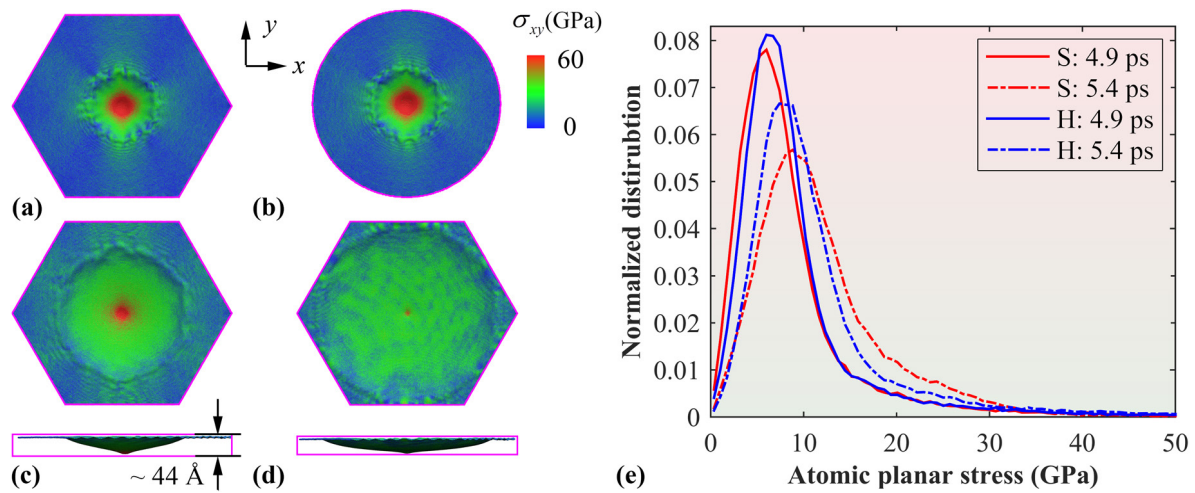

**Figure 5** Planar stress ($\sigma_{xy}$) distribution for: (a) hexagonal graphene membrane at 5.4 ps, (b) circular graphene membrane at 5.4 ps, (c) hexagonal graphene membrane at 8.0 ps, and (d) hexagonal graphene membrane at 10.4 ps; (e) Distribution function of planar stress for square and hexagonal graphene at different simulation time. S and H represent the square and hexagonal graphene membrane, respectively.

*Failure shape*

Another important characteristic with the impact process is the failure shape of graphene after penetration. Recall Figure 3e and 4c, the graphene membrane exhibits different failure shapes when the boundary condition changes. Thus, following efforts will try to find out the general failure shape of monolayer graphene. Considering that the ratio between the projectile diameter($D$, ~ 6 nm) and the thickness of graphene ($h$, ~ 3.35 Å) is much larger than one, i.e., $D/h \gg 1$, the penetration energy can be estimated from[14]

$$E_P = (\rho A_s h) v^2 / 2 + E_d \qquad (2)$$

Here, $\rho$ is the mass density, $A_s$ is the strike face area ($A_s = \pi r^2$, $r$ is the radius of the projectile), and $v$ is the impact velocity. The first term represents the minimum inelastic energy transferred to the sample, and the second term $E_d$ represents all other energy dissipation during impact (e.g., elastic deformation of the sample, and air drag – ignored in the simulation). According to Eq. 2, for a given sample, the penetration energy is related to projectile velocity and radius. In this regard, we first consider how the projectile radius would influence the failure shape of the graphene membrane. Same as above, a fully-fixed 50 × 50 nm² square graphene membrane is considered. Projectile with a same impact velocity of 20 Å/ps but different radius ranging from 24 to 34 Å is adopted. As expected, the penetration energy increases with the projectile radius (Figure 6a), e.g., $E_p$ is around 4156 eV for the largest examined projectile (34 Å), which is over two times higher than the results obtained from the projectile with a radius of 25 Å (~ 1745 eV). When its radius is reduced to 24 Å, the projectile is bounced back with no penetration occurrence observed.

As expected, with changing impact velocity, the failure shape of the graphene varies (Figure 6b to 6e). Specifically, with increasing impact velocity, the edge number of the crack polygons increases from three to six, i.e., the crack number increases from three to six. Similar as observed from previous cases, all crack edges have zigzag edges and they all align with the six potential crack directions. For simplicity of discussion, we denote these directions as three direction pairs (*a*, *a'*), (*b*, *b'*), and (*c*, *c'*), where the two directions in each pair are the inverse direction to each other (Figure 6b), and they have fixed angles with the *x*-axis as 0°, 60° and 120°, respectively. Therefore, based on the crack polygons, we can measure the total crack length ($L_{crack}$) according to

$$L_{crack} = \sum_i^n (L_i^{ideal} - L_i^{melt}) \qquad (3)$$

where $L_i^{ideal}$ is the ideal length of the *i*th crack, and *n* is the number of cracks (only major cracks that form the final crack shape is considered). $L_i^{melt}$ is stemmed from the fact that the immediate contact area of the graphene might melt due to high impact energy in some cases (especially under high impact velocity as discussed in the following), thus, the effective crack length should subtract this part if melt is observed. Assuming that the melt area is a circular region with the same radius as the projectile, its edge profile can be simply expressed as $(x-x_0)^2 + (y-y_0)^2 = r^2$, with $(x_0, y_0)$ as the projectile centre coordinate. Thus, depending on the pairs that the crack belongs to, its ideal and melt length can be easily calculated based on the coordinates of the crack tip $(x_1, y_1)$ and the intersection $(x_2, y_2)$ between the crack line and the melt edge according to (see Figure 6d)

$$\left. \begin{array}{ll} L_{ideal} = |x_1 - x_0|, & L_{melt} = |x_2 - x_0| \quad (a,a') \\ L_{ideal} = |y_1 - y_0|/\sin(\pi/3) & L_{melt} = |y_2 - y_0|/\sin(\pi/3) \quad (b,b') \\ L_{ideal} = |y_1 - y_0|/\sin(\pi/3) & L_{melt} = |y_2 - y_0|/\sin(\pi/3) \quad (c,c') \end{array} \right\} \quad (4)$$

To ensure a good estimation, Eq. 4 has used the original coordinates of the atoms (before impact) to avoid the influence from the deflection of the graphene petals during impact. Based on Eq. 3 and 4, the estimated total crack length is found to increase with the projectile radius, which shares the same changing tendency with that of the penetration energy (Figure 6a). For instance, a total crack length of about 1285 Å is estimated for the projectile radius of 34 Å, which is over two times longer than that induced by the projectile radius of 25 Å ($L_{crack}$ ~ 506 Å). It is assumed that larger projectile will induce more significant local deformation to graphene before failure, i.e., require higher penetration energy. After penetration, the stored strain energy is released and cause more bonds to break. Essentially, besides the energy loss in the impact area, majority of the penetration energy is consumed by the crack growth. Therefore, the total crack length is supposed to have a strong relationship with the penetration energy, which is seen from the simulation results in Figure 6a and also following results in Figure 7a.

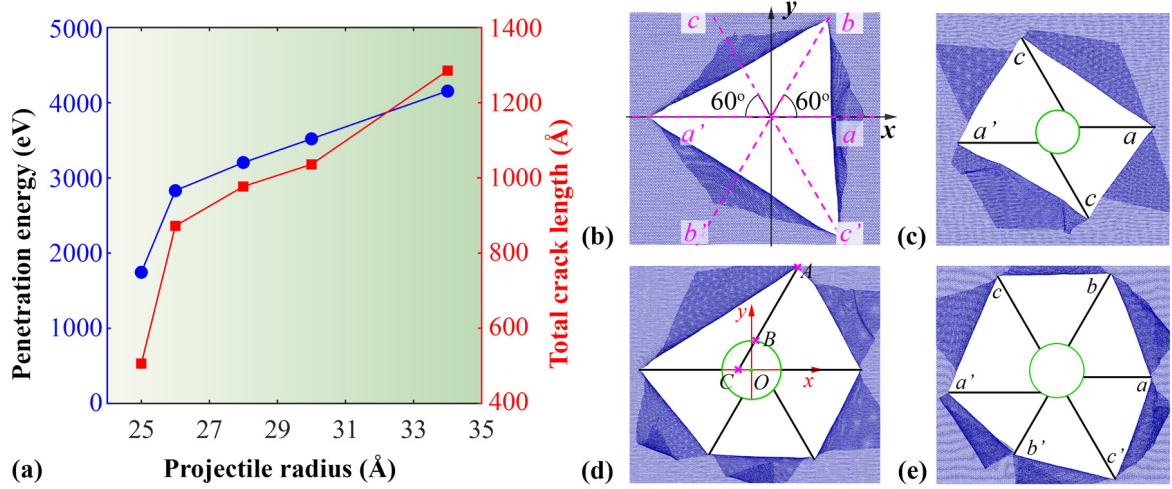

**Figure 6** (a) Penetration energy and the total crack length as a function of projectile radius; Atomic configurations of the graphene membrane showing the final failure shape induced by the projectile with a radius of: (b) 25 Å, (c) 26 Å, (d) 28 Å, and (e) 34 Å. Figure b shows the six preferential crack directions denoted by (*a*, *a'*), (*b*, *b'*) and (*c*, *c'*). Figure d schematically illustrates the geometrical relations between the crack tip *A* ($x_1$, $y_1$), and the intersections *B* ($x_2$, $y_2$) and C. Here, distance AC is defined as the ideal crack length, and BC is referred as the melt crack length. Solid dark lines in Figure c, d, and e schematically show the cracks formed after impact, and the green circles represent the projectile. To note that the atomic configurations have been scaled to a similar size, which do not reflect the actual size of the sample and the projectile.

We then consider the failure shape of graphene under different impact velocity values, ranging from 20 to 60 Å/ps. As shown in Figure 7a, the penetration energy first decreases and then increases with the velocity (for velocity higher than 3 km/s). The increasing portion exhibits a parabolic relationship with the velocity ($\propto v^2/2$), in line with the previous supersonic projectile penetration studies.[14,15] Similarly, we found a same changing tendency for the estimated total crack length, and the crack number increases with the increases of impact velocity. For instance, three major cracks are found at the impact velocity of 20 Å/ps, and six cracks (hexagonal shape) are formed under the velocity over 40 Å/ps. The decreasing trend of the penetration energy/crack length can be explained according to the fact that the graphene will experience more elastic deformation (before failure) at low impact velocity, which is analogue to an indentation deformation.[2,34]

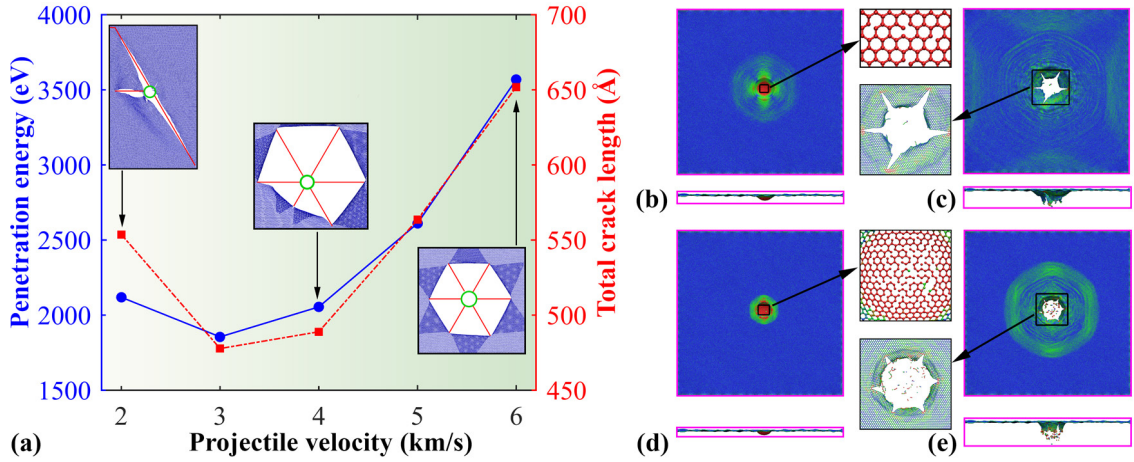

**Figure 7** (a) Penetration energy and the total crack length as a function of the projectile velocity. Insets schematically show the atomic configurations of the failure shape at the velocity of 2, 4 and 6 km/s; Planar stress ($\sigma_{xy}$) distribution for the graphene membrane under projectile velocity of 3 km/s at: (b) 2.6 ps, and (c) 3.6 ps; Planar stress ($\sigma_{xy}$) distribution for the graphene membrane under projectile velocity of 6 km/s at: (d) 1.2 ps, and (e) 1.8 ps;

To further reveal the velocity impacts, we compared the atomic configurations of the graphene under the velocity of 30 and 60 Å/ps. As is seen, the graphene experiences extensive elastic deformation before the onset of crack generation and propagation under 30 Å/ps. However, for the 60 Å/ps case, the impact area is melted immediately when the projectile approaches graphene, which creates a large number of dangling bonds (inset of Figure 7d). For both cases, a clear tensile stress wave surrounding the impact area is found (Figure 7b and 7d), which spreads quickly to rest of the sample. After penetration, the release of the energy stored in the petals leads to the continuing growth of cracks, which eventually form a pentagon and hexagon (inset of Figure 7a) failure shape, respectively.

It is crucial to mention that the creation of dangling bonds (especially at the initial impact stage) is due to the hypervelocity impact, and the time step (i.e., 0.5 fs) adopted here is accurate to reproduce this phenomenon. From the impact tests under a high impact velocity of 6 km/s, the number of breaking bonds in graphene under different time steps (ranging from 0.05 to 1 fs) agrees well with each other. As illustrated in Figure 8a, the number of breaking bonds in all cases follows a similar increasing pattern and range. From the energy curves, the estimated penetration energy fluctuates around 3230 eV with a small standard deviation of 56 eV, and no clear relationship is detected between the penetration energy and the examined time step range (see Supporting Information). As also reflected from the atomic configurations (insets of Figure 8a), although the fracture pattern shows certain difference while applying different time steps, the number of breaking bonds are essentially similar and comparable. Similar results have also been observed from the impact tests under a smaller impact velocity of 3 km/s, and the penetration energy is found to fluctuate around 1752 ± 38 eV (see Supporting Information). These results signify that the time step as adopted in this work is valid for the hypervelocity impact simulation purpose.

One striking feature between Figure 7c and 7e is the clear difference in the tensile stress wave pattern at same penetration distance. For Figure 7c (from the projectile velocity of 3 km/s), the stress wave has already bounced back by the boundary, whereas, the stress wave is still transmitting to the boundary for graphene under the velocity of 6 km/s (Figure 7e). To analysis the influence on the crack growth from the stress wave, we tracked the crack growth rate for the graphene. In this regard, we compared the results from the square graphene membrane under impact velocity of 6 km/s with a side length of 50 nm and 100 nm. Since all six cracks in these two cases are almost identical to each other (inset of Figure 7a), we only track the crack in direction $c$. As illustrated in Figure 8b, the crack length for both samples increases continuously with simulation time and its profile can be well described by a power-law relationship. Specifically, the crack is found to propagate until ~ 20.5 ps for the smaller sample (50 nm), whereas, it stopped propagation at around 16.4 ps for the larger sample (100 nm). Based on the fitted power-law relationship, the corresponding crack growth velocity can be derived. As shown in Figure 8b, the crack growth velocity shows a decreasing trend, which is understandable as the aggregated strain energy reduces with the crack growth.

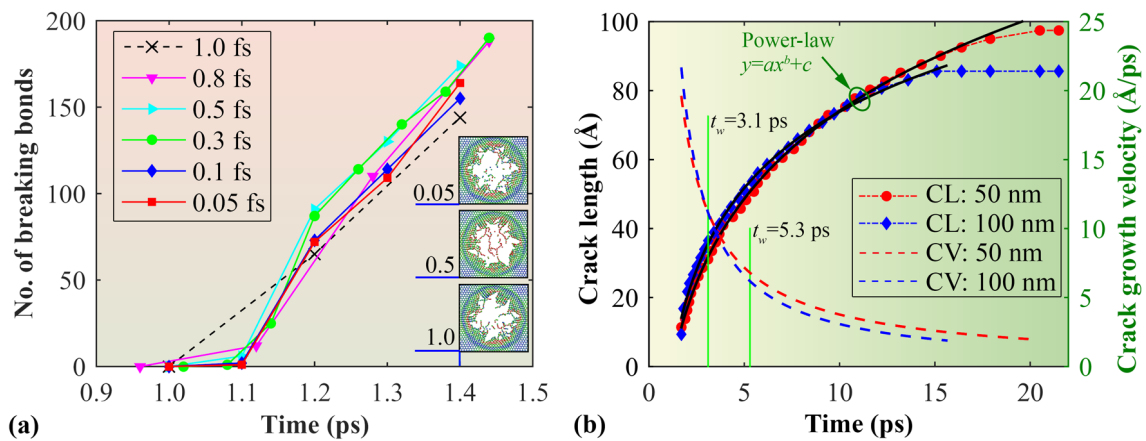

**Figure 8** (a) Number of breaking bonds in the graphene membrane as a function of time during the initial penetration under different time steps (with a projectile velocity of 6 km/s, sample size 50 × 50 nm$^2$). Insets show the atomic configurations of the graphene membrane (in the penetration region) at the simulation time of 1.4 ps under the time step of 0.05, 0.5 and 1.0 fs, respectively. (b) Crack length (in $c$ direction) and crack growth velocity as a function of time for the graphene membrane under a projectile velocity of 6 km/s. CL and CV represent crack length and crack growth velocity, respectively. Here, the crack length is estimated by tracking the number of breaking bonds during the whole simulation. The two solid green lines highlight the approximate time when the first stress wave bounced back to the impact centre.

It is noted that although the larger sample shows slightly higher penetration energy (~ 3608 eV) than the smaller sample (~ 3568 eV), its crack in *c* direction is ~ 85.6 Å, which is shorter comparing with the smaller sample (~ 97.5 Å. The total crack length also shows similar phenomenon. Such result could be explained from the perspective of the tensile stress wave bounced back by the fixed boundary, which is supposed to help the crack growth. Theoretically, the tensile stress wave in monolayer graphene has an ultra-high speed of 22.2 km/s (222 Å/ps). In other words, it only takes around 1.1 ps for the stress wave to reach the boundary for the square graphene membrane (size of 50 nm). Considering the simulation time before the projectile approaching the sample (~ 0.9 ps) under the velocity of 6 km/s, the stress wave would bounce back to the impact area at approximately 3.1 ps for the 50 nm graphene, and ~ 5.3 ps for the 100 nm graphene (highlighted in Figure 8b). That is, the returning stress wave will affect the crack length much earlier in smaller sample. Moreover, the stress wave generated by the impact is a typical Rayleigh wave,[35] and its amplitude in the out-of-plane direction decays as $1/\sqrt{R}$ (where $R$ is the radial distance from the impact centre), that is, its energy decays with travelling distance. Therefore, for the lager sample, the returning stress wave also possesses much less energy, and thus exerts less impact to the crack growth. Overall, increase the graphene membrane size will weaken the influence on the crack growth from the stress wave bounced back by the boundaries.

Above results have suggested that the failure shape of graphene will change with projectile radius and impact velocity. Recall the boundary influence on stress distribution and transmission during impact, here we conduct another four groups of simulations. Particularly, we revisited the square graphene with two fixed boundary, the hexagonal graphene, and the circular graphene, by applying various impact velocity ranging from 20 to 60 Å/ps. At the lowest impact velocity (20 Å/ps), no penetration is observed for the hexagonal and circular graphene membrane. As compared in Figure 9a, the specific penetration energy in each group follows the general parabolic relation with the impact velocity as given by Eq. 2. Here, the specific penetration energy $E_p{}^*$ is defined as $E_p{}^* = E_p/(\rho A_s h)$. Meanwhile, we also found obvious deviations between the parabolic baseline and the estimated penetration energy, which is supposed as induced by the specific delocalized penetration energy. At different impact velocity, we expect different amount of penetration energy being delocalized. Agree with our previous discussion, circular graphene membrane generally absorbs higher specific penetration energy, followed by hexagonal graphene. It is worth mentioning that by normalizing the specific penetration energy, the influence from the fixed-boundary condition appears similar under different impact velocity values (see discussions in Supporting Information). That is, the fixed-boundary condition exerts insignificant influence to the penetration energy under different impact velocities as examined in this work.

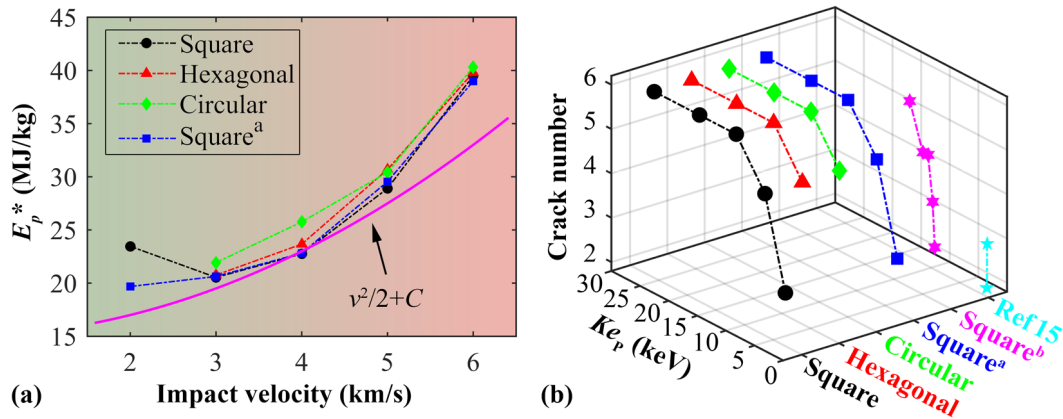

**Figure 9** (a) Specific penetration energy as a function of the impact velocity for different graphene membrane models. Square[a] represents the square graphene membrane with two fixed boundaries, other models are fully fixed; (b) The total crack number as a function of the initial kinetic energy of the projectile from different examined structures. Square[b] represents the results from the testing with same impact velocity (2 km/s) but different projectile radius (from 24 to 34 Å).

Before conclusion, we summarised the failure shape (the crack number) as obtained from different testings. As compared in Figure 9b, the crack number has a clear relationship with the kinetic energy of the projectile ($Ke_p$), i.e., it increases monotonically with the kinetic energy. Specifically, for $Ke_p$ higher than ~ 7 keV, the graphene shows six cracks (i.e., a hexagonal failure shape). For $Ke_p$ between ~ 4 and 7 keV, the graphene exhibits five cracks, and the crack number decreases further if the projectile has smaller kinetic energy. These results also explain why the graphene only exhibits two or three cracks in the work by Haque et al[16] (as the projectile has a relatively small kinetic energy, less than 300 eV).

**Conclusions**

Based on large-scale molecular dynamics simulations, we explored the fracture behaviour of monolayer graphene under hypervelocity impact. It is found that the cracks are formed preferentially in the zigzag directions which are consistent with that observed from tensile deformation. Specifically, the boundary condition is found to exert an obvious influence on the stress distribution and transmission during the impact process, which eventually influences the penetration energy and crack growth. For similar sample size, the circular shape graphene possesses the best impact resistance, followed by hexagonal graphene membrane. Moreover, it is found the failure shape of graphene membrane has a strong relationship with the initial kinetic energy of the projectile. The higher kinetic energy, the more number the cracks. For instance, for initial projectile kinetic energy higher than around 7 keV, the graphene membrane will show a hexagonal failure shape (with six cracks). In comparison, fewer cracks will appear when the kinetic energy decreases. This study provides a fundamental understanding of the deformation mechanisms of monolayer graphene under hypervelocity impact, which should shed lights on the design of graphene-based textile for bullet-proof application, or shielding structure for aerospace systems. We note that although these results are obtained from monolayer graphene under a low temperature, similar failure mechanisms are also expected under higher temperature or multi-layer graphene membrane, which still deserves further investigation.

**AUTHOR INFORMATION**

**Supporting Information**

Supporting information is available for the further discussion on the influence from time step, fixed-boundary condition, and sample size.

**Corresponding Author**

zhan.haifei@qut.edu.au; yuantong.gu@qut.edu.au

**Acknowledgments**

Supports from the ARC Discovery Project (DP130102120), the High Performance Computer resources provided by the Queensland University of Technology, the National Natural Science Foundation of China (11272118) are gratefully acknowledged.

**Author contributions**

K. Xia performed the simulation. K. Xia and H.F. Zhan analysed the data and wrote the paper. D.A. Hu and Y.T. Gu revised the paper. K. Xia, H.F. Zhan, D.A. Hu and Y.T. Gu discussed the results and analysis.

**Additional information**

Competing financial interests: The authors declare no competing financial interests.